\documentclass[conference, final]{IEEEtran}
\IEEEoverridecommandlockouts
\usepackage{cite}
\usepackage{amsmath,amssymb,amsfonts}
\usepackage{algorithmic}
\usepackage{graphicx}
\usepackage{textcomp}
\usepackage{xcolor}
\def\BibTeX{{\rm B\kern-.05em{\sc i\kern-.025em b}\kern-.08em
    T\kern-.1667em\lower.7ex\hbox{E}\kern-.125emX}}

\begin{document}

\title{Recent Advancements in Microscopy Image Enhancement using Deep Learning: A Survey}

\author{\IEEEauthorblockN{1\textsuperscript{st} Debasish Dutta}
\IEEEauthorblockA{ \hspace{4em} \textit{Dept. of Computer Science \hspace{4em}} \\
\textit{Gauhati University}\\
Assam, India \\
debasish@gauhati.ac.in}
\and
\IEEEauthorblockN{2\textsuperscript{nd} Neeharika Sonowal}
\IEEEauthorblockA{ \hspace{7em} \textit{Dept. of Computer Science \hspace{7em}} \\
\textit{Gauhati University}\\
Assam, India \\
neeharika@gauhati.ac.in}
\and
\IEEEauthorblockN{3\textsuperscript{rd} Risheraj Barauh}
\IEEEauthorblockA{ \hspace{2em} \textit{Dept. of Computer Science}\hspace{2em} \\
\textit{Gauhati University}\\
Assam, India \\
rishirajbaruah@gauhati.ac.in}
\and
\IEEEauthorblockN{4\textsuperscript{th} Deepjyoti Chetia}
\IEEEauthorblockA{\hspace{2em}\textit{Dept. of Computer Science}\hspace{2em} \\
\textit{Gauhati University}\\
Assam, India \\
deepjyotichetia@gauhati.ac.in}
\and
\IEEEauthorblockN{5\textsuperscript{th} Dr. Sanjib Kr Kalita }
\IEEEauthorblockA{ \hspace{1em} \textit{Dept. of Computer Science}\hspace{2em} \\
\textit{Gauhati University}\\
Assam, India \\
sanjib959@gauhati.ac.in}
}

\maketitle

\begin{abstract}
Microscopy image enhancement plays a pivotal role in understanding the details of biological cells and materials at microscopic scales.
In recent years, there has been a significant rise in the advancement of microscopy image enhancement, specifically with the help of deep learning methods. This survey paper aims to provide a snapshot of this rapidly growing state-of-the-art method, focusing on its evolution, applications, challenges, and future directions. The core discussions take place around the key domains of microscopy image enhancement of super-resolution, reconstruction, and denoising, with each domain explored in terms of its current trends and their practical utility of deep learning.

\end{abstract}


\section{Introduction}
\begin{figure}[htbp]
    \centerline{\includegraphics[width=0.94\linewidth]{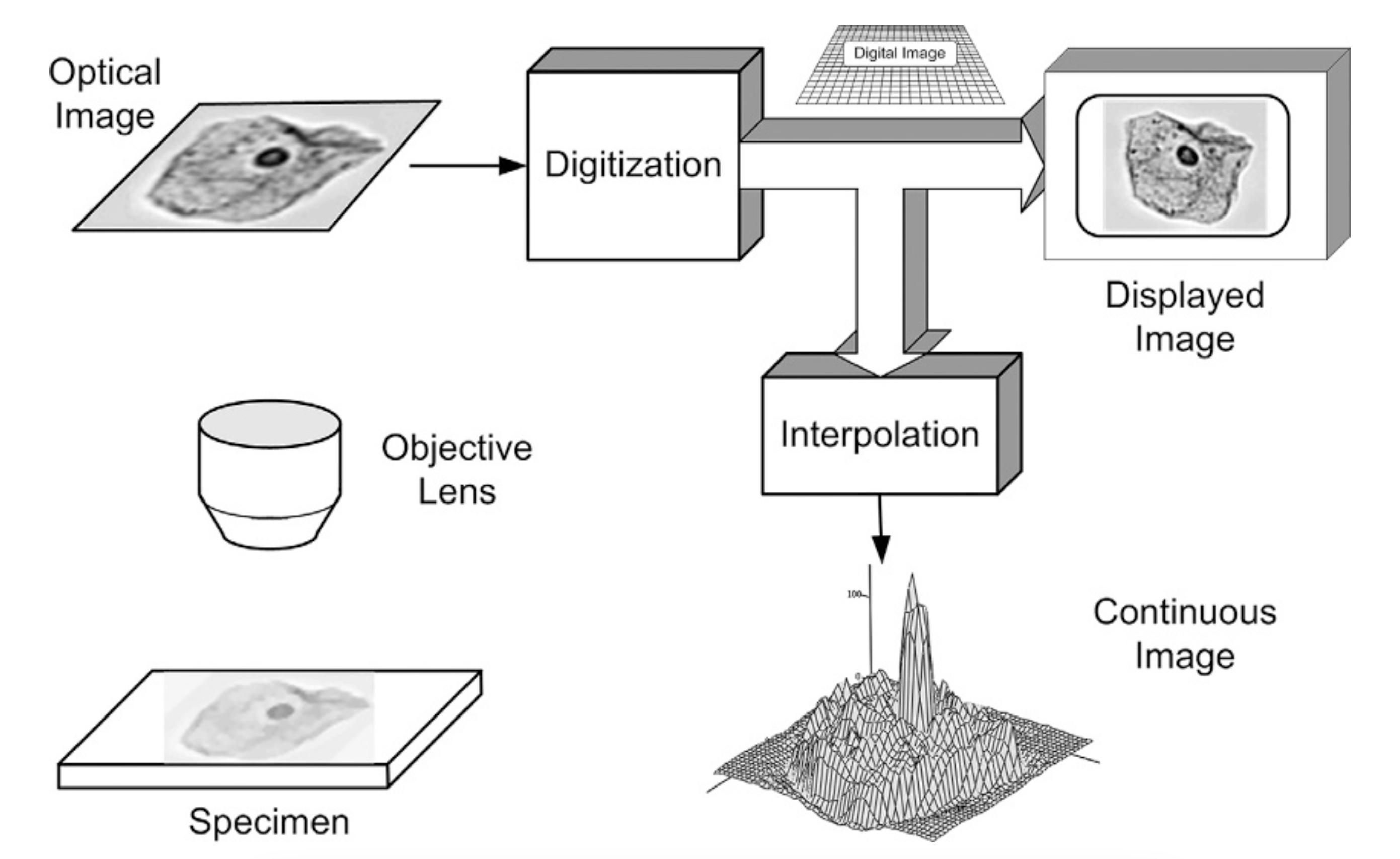}}
    \caption{Image Acquisition Process in a Microscope.}
    \label{figmicros}
\end{figure}

Microscopy plays a crucial role in various scientific and medical disciplines, enabling researchers to explore the intricate details of biological specimens, materials, and structures at the micro and nanoscales. Four types of images occur in a traditional microscopy process - there is the \emph{optical image}, which gets captured by the lens, and then forms the \emph{digital image} after essential digitization. After applying image preprocessing techniques, it is developed into either the \emph{displayed image} consisting of just the two dimensions  ($x\ width\ and\ y\ height$ ) for screens or a \emph{continuous image} (depending on the type can be up to 5-dims including $z\ depth$, $\lambda\ wavelength$ and $t\ time$ ), as depicted in ``Fig.~\ref{figmicros}''. Most microscopy imaging techniques are applied while or after digitizing to avail maximum benefits. Due to the limited capacity of optical imaging with the combination of inherent noise, image enhancement techniques are much more favorable in almost all applications, thus leading to the rapid development of microscopy image enhancement methods and techniques.

The paper is organized into two major sections: the background, which briefly delves into the conventional Microscopy image enhancement techniques and the need for deep learning (DL) in the field, and then, the discussion section, which is categorized into three groups according to the primary task they handle. For each part, the first the recent methods are discussed, followed by detailed applications of the methods, and finally, conclude with the challenges and development of each task.

\section{Background}

\subsection{Traditional techniques in Microscopy Image Enhancement}
One of the most essential tools used in biological sciences for live cells and tissues is Fluorescence microscopy (FlM). It is an optical microscopy modality that is very widely used. However, as with all optical modalities, it suffers from low-resolution limits due to high light diffraction. Then there is the Electron microscopy (EM) technique, which obtains high-resolution images of biological and non-biological specimens. However, a significant disadvantage is its inability to work with live cells and organisms, thus the need to fall back towards optical microscopes. Thus, image enhancement, especially in optical microscopes, became pretty much a necessity.

\begin{figure}[htbp]
    \centerline{\includegraphics[width=\linewidth]{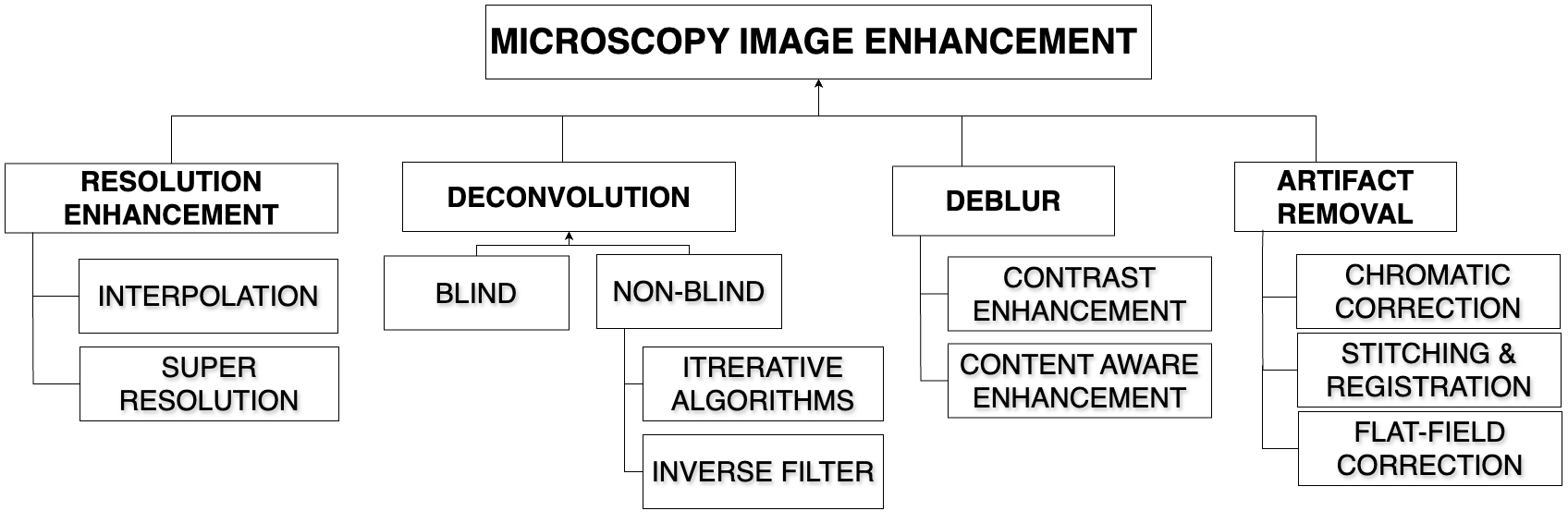}}
    \caption{Taxonomy of Microscopy Image Enhancement Techniques.}
    \label{figTaxo}
\end{figure}

Before the advent of deep learning for microscopy image enhancements, it has been done through manual computational methods. The various algorithms developed for microscopy applications are classified into two broad categories - the spatial domain and the transfer domain\cite{Castleman2008}. Based on image statistics, the techniques and algorithms that operate on the whole or part of the image fall under the spatial domains, such as histogram sharpening. Those enhancement methods manipulating images through alternative domains achieved after a mathematical operation, such as Fourier, Sparse or Wavelet transform, are categorized into transfer domains. A detailed classification of the techniques are shown in the in ``Fig.~\ref{figTaxo}''.

For a gray image with an intensity range $[0,L]$,
\begin{equation}
    g=T(I)\label{eqGG}
\end{equation}
a \textit{global operation} of the image $T$, will maps the image $I$ to a new image $g$, according to ``\eqref{eqGG}''.
Any operation under such form is considered to be in the spatial domain. These include high, low, band-pass filters, denoising operations such as median filters and Gaussian noise filters, deconvolution algorithms such as Richardson–Lucy (RL) or maximum likelihood estimation (MLE) deconvolution, resolution enhancement techniques like bicubic and Lanczos interpolation, and flat-field corrections for artifact removal.

However, it becomes challenging to work directly on the spatial features of the images, and this is where the spatial domain is converted into a different domain like the Fourier transform is used to convert it into the Fourier domain, which makes calculations much easier\cite{beghdadi1989}. Deconvolution methods such as Wiener filters and low and high-pass filters use the Fourier domain. Also, there is the Wavelet Transform\cite{knutsson1983} , which dissects the image signals into frequency components across different scales, making it more suitable for applications such as noise reduction and edge detection. Thus, wavelet thresholding is a significant enhancement technique for noise removal. Such techniques are collectively called to be in the transfer domain.

\subsection{Need for Deep Learning in Microscopy Image Enhancement}
With the advent of Deep Neural Networks (DNNs) in the last decade, works such as \cite{krizhevsky2017} and \cite{simonyan2015} led deep learning to become a significant part of various image processing tasks, including but not limited to image enhancement. Thus, image enhancement in microscopy has also advanced in recent years. Applications which were previously limited due to the computational capabilities are unlocked, allowing researchers to explore new horizons. For tasks such as denoising, deconvolution, resolution enhancement, and artifact removal, when done using DL, the results are far superior compared to previous works using traditional filters. Jin\cite{jin2020} showed the use of DL to produce high-quality Structured illumination microscopy (SIM) images with very few low-quality images as input, which previously needed large amounts of very high-quality images. They also showed how DL could effectively be used for cell segmentation, denoising, and super-resolution. Also, DL can be used in image restoration through content awareness, which has produced excellent results in denoising superior signal-to-noise ration (SNR). De Haan\cite{dehaan2020} argued how DL-based image enhancement and reconstruction, especially regarding optical microscopy, has mainstream applications. They discussed how DL can solve inverse problems in microscopy, like deconvolution and denoising across different modalities. They also realized that doing such tasks using DL is much more cost-effective. Deep learning allowed optical microscopes to capture images on par with those of EM ones. Pradhan \cite{pradhan2020} discussed how deep learning became a massive boon for biophotonics, including resolution enhancement and pseudo-staining. They also discussed the potential use of DL and applications for biophotonic data. Using generative networks for resolution enhancement and denoising is much more effective than conventional methods such as the Wiener filters or Richardson-Lucy algorithms. Super-resolution methods employed through DNNs can even surpass the diffractive limit of the microscope to get a much higher-quality image. Also, the enhanced microscopic image is much more effective for further tasks like analysis and localized segmentation. Kwon \cite{kwon2019}proposed a super-resolution enhancement of an Integral Imaging Microscopy(IIM) system using dl, which has proven to be highly effective in improving the resolution and can use the optimized images as learning data further down the line.
Overall, this need for dl in microscopic image enhancement arises from the ability to surpass traditional quality, efficiency, and applicability methods across various modalities. Deep learning techniques offer a cost-effective and powerful solution to challenges like denoising, resolution enhancement, artifact removal, and image quality improvement in microscopy.

\subsection{Development of Deep Learning methods for Microscopy Image Enhancement}

For image processing using DL, convolutional neural networks (CNN)\cite{simonyan2015} are immensely popular as they give excellent results because they work on each pixel by assigning a label to it, thus enabling pixel-wise classification. Hence, they work excellently as feature selectors, while traditional fully connected neural layers form their last layers. Therefore, even while designing generative adversarial networks (GAN)\cite{gan}, CNNs are used for the generator part. Gradually, all these CNN and GAN models are being applied to the field of microscopy image analysis.

\begin{table*}[htbp]
    \caption{Summary of some key deep learning application in Microscopy image Enhancement}
    \begin{center}
    \begin{tabular}{|c|c|c|c|c|c|}
    \hline
    \textbf{Network}& \textbf{Year} & \textbf{Task} & \textbf{Architecture} & \textbf{Results} \\
    \hline
    \hline
    cGAN & 2023 & SR & TCAN\cite{huang2023a} & MSE=25 \& PSNR=13 \\ \hline
    GAN & 2023 & SR & Real-ESRGAN\cite{dong2023} & \\ \hline
    CNN + Attn & 2022 & SR & SF-SIM\cite{cheng2022} & PSNR=31.19 \& SSIM=0.732 \\ \hline
    GAN & 2021 & SR & IIM-GAN\cite{alam2021} & PSNR=37.84 \& SSIM=0.99 \\ \hline
    GAN & 2021 & SR & ESRGAN*\cite{dai2021} & PSNR=35.85 \& SSIM=0.85 \\ \hline
    U-Net & 2020 & SR & UNet-SIM15, ScUNet\cite{jin2020} & PSNR=20.32 \& SSIM=0.40 \\ \hline
    VGG & 2019 & SR & SR-NET\cite{tom2019} & PSNR=43.04 \& SSIM =0.97 \\ \hline
    UNET & 2023 & Restoration & RedrawNet\cite{han2023} & Accuracy: 0.9086 \\ \hline
    UNet-GAN & 2022 & Restoration & AR-PAM\cite{zhang2022c} & PSNR=24.39 \& SSIM=0.617 \\ \hline
    TL & 2023 & Deconv & CiDeR\cite{chobola2023} & PSNR=30.63 \& SSIM=0.8925 \\ \hline
    CNN & 2022 & Deconv & PhaseNet\cite{vinogradova2022} & MSE=0.001554 \\ \hline
    GAN & 2020 & Deconv & CycleGAN with blur kernel\cite{lim2020} & PSNR=26.62 \& SSIM=0.9573 \\ \hline
    GAN & 2019 & Deconv & SpCyGAN\cite{lee2019} & M-IFQ=0.52 \& 0.92 \\ \hline
    CNN & 2019 & Deconv \& Denoise & CBS-Deep\cite{bai2019} & PSNR=27.91 \\ \hline
    ResNet & 2022 & Deblur & ResUNet\cite{sanghvi2022a} & PSNR=24.20 \\ \hline
    CNN & 2020 & Deblur & RDN\cite{zhao2020} & PSNR=21.039 \& SSIM=0.8507 \\ \hline
    VGG16 & 2020 & Deblur & DBMID\cite{jiang2020} & SSIM=0.78 \\ \hline
    TL & 2022 & Denoise & DnCNN\cite{demircan-tureyen2022} & PSNR=37.01 \& SSIM=0.924 \\ \hline
    GAN & 2022 & Denoise & MI3DA\cite{fuentes-hurtado2022} & PSNR=37.07 \& SSIM=0.95 \\ \hline
    DCNN & 2021 & Denoise & BoostNET\cite{vo2021} & PSNR=35.62 \& SSIM=0.9129 \\ \hline
    Encoder$/$ Decoder & 2021 & Denoise & IRUNET\cite{gilzuluaga2021} & PSNR=38.38 \& SSIM=0.98 \\ \hline
    CNN & 2019 & Denoise & Noise2Void\cite{krull2019} & PSNR=~32.28 \\
    \hline
    \end{tabular}
    \label{tabDl}
    \end{center}
    \end{table*}

\section{Discussions}

\begin{figure}[htbp]
    \centerline{\includegraphics[width=\linewidth]{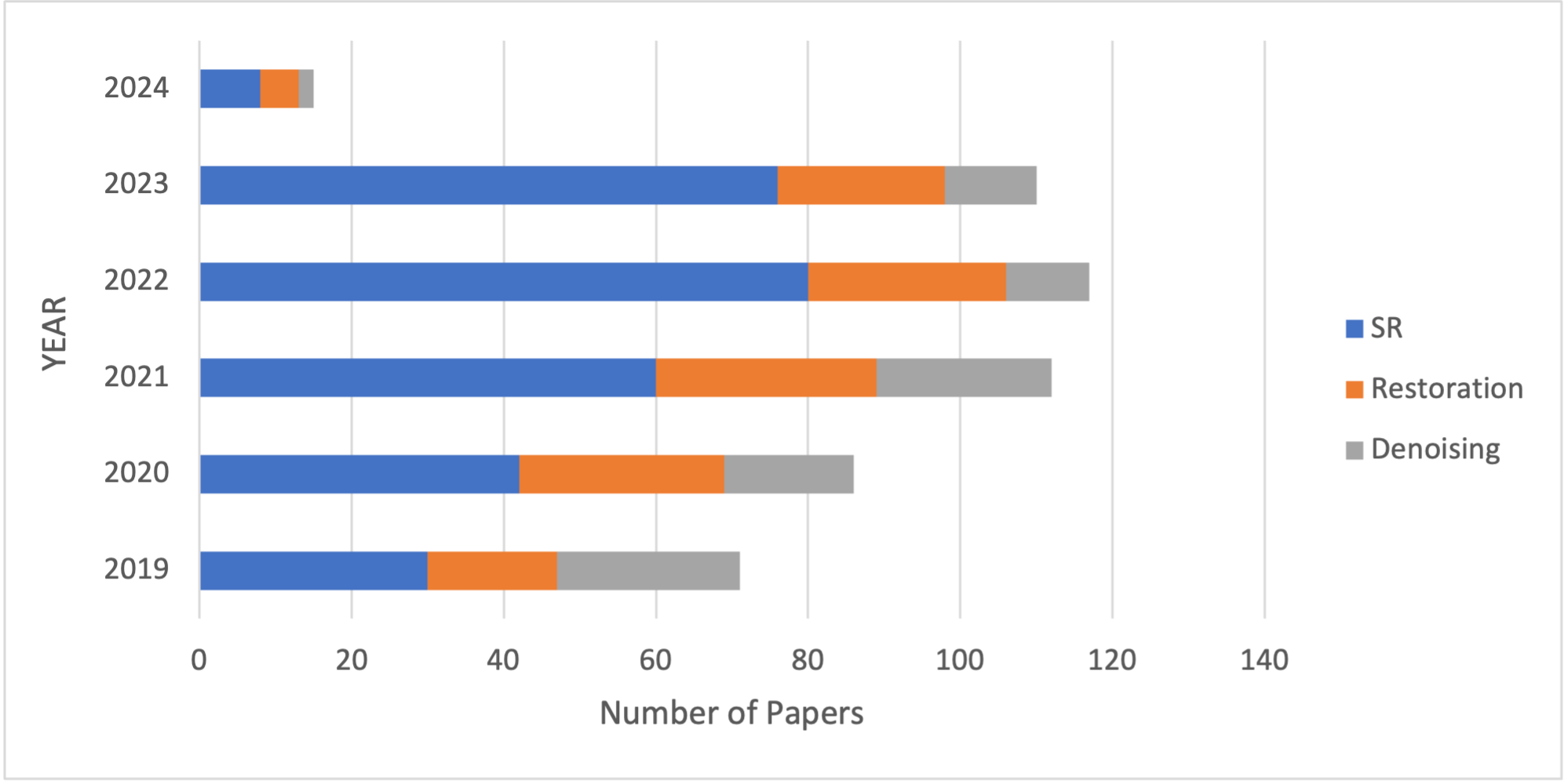}}
    \caption{Number of papers in last five years.}
    \label{figGraph}
\end{figure}

In this study, last five years notable work in this field of microscopy image enhancement have been analyzed and studied. Although enhancement is done in various fields, it has been categorized into three major parts: resolution enhancement through SR, image restoration using deconvolution and deblurring and image denoising. A graph depicting the number of papers published each year arranged with their categories is shown `Fig~\ref{figGraph}'.

An analysis of most recent works are shown in `Table~\ref{tabDl}'.

\subsection{Super Resolution}
Super-resolution (SR) started with restoring missing pixels after capturing, followed by resolution enhancement achieved through spatial and transform domain methods. But now, after the arrival of super-resolution, its objective has evolved to increase the image's resolution beyond that of the original, thus overcoming the resolution limit of the capture device itself.

Adversarial networks have been gaining fame, primarily due to their superior performance in improving simple image resolution. Using deep learning in super-resolution eliminates the requirement of numerical modeling of the imaging process or estimating a point-spread function\cite{Wang2018}. Deep learning techniques like GANs are used to transform diffraction-limited input images into super-resolved ones.

Tom et al. introduced a Super-Resolution Network (SRNet)\cite{tom2019} model using deep convolutions that minimize the reconstruction loss between the real and enhanced images and maximize the discrimination done by relativistic visual turing test (rVTT) networks to achieve good results. Due to the introduction of rVTT as a discriminator, the loss of perceptual index is significantly lower during training, and the results are evident in their high peak signal-to-noise ratio (PSNR) and structural similarity index measure (SSIM) values.

SIM is a fluorescence microscopy technique used for wide-field microscopy that uses light patterns to excite the sample, making them a standard for a large field-of-view, long term imaging and SR.
Although SIM offers to surpass optical diffraction limit and produce better resolution over conventional optical microscopy, it requires multiple high-resolution (HR) images of the same sample along with explosive illumination. So, using DL, Jin L et al. \cite{jin2020} proposed a method that gives an over a 5-fold reduction in the number of HR images required and under extremely low light for an SR-based SIM. It uses a UNet\cite{unet2015} as the GAN and only requires 50–70 samples per structure and training for 2000 epochs.
Alam et al.\cite{alam2021} created an SR enhancement method for Integral-imaging microscopy (IIM) using a GAN. Their GAN solves singke-image SR  in a two-step process. Before feeding the images to the GAN, a microlens array (MLA) is used to capture the IIM images. The GAN's generator consists of a residual block and an upsampling block with parametric ReLU, while the discriminator used the classic C-B-L (convolution - batch normalization - leaky ReLu layers) to create and distinguish between HR and low-resolution (LR) images. This simple yet intuitive method does not require any iterative interpolation-based images, and its generation time reduces each iteration.
Furthermore, many of the state-of-the-art (SOTA) algorithms are involved in developing SR images in microscopy use ESRGAN (Enhanced Super-Resolution Generative Adversarial Network)\cite{wang2019b} as a base point due to its excellent results in generating high-resolution images. Dai M et al.\cite{dai2021} demonstrated how using DL,  resolution can be enhanced for regular and mini microscopy. Their method adopted ESRGAN and built upon it by updating the original parameters using a network interpolation strategy during testing. This modified ESRGAN provided artifact-free, higher reconstruction accuracy and better supervision for texture recovery. Dong S et al.\cite{dong2023} then introduced a method to enhance the resolution of images in digital holographic microscopy (DHM) by using ESRGAN along with a low NA objective lens (Real-ESRGAN). This method can take the images captured using a low NA lens and, after feeding it to their R-ESRGAN, can give images comparable to those captured by a high NA lens. They used a Residual-in-residual dense block (RRDB), which helps recover the details and structure of HR images by training the generator. Thus, this method has the potential to significantly reduce the cost of DHM imaging, making it a promising technique with a wide variety of applications. Also, self-attention-based mechanisms have been getting quite popular, as Cheng et al.\cite{cheng2022}introduced a band pass attention block, added to their Fast and lightweight Network (FLSN) along with a noise estimator module. This allowed them to use just a single frame in SIM to beat previous algorithms that use up to 15 frames and also in many adverse conditions.

\subsection{Reconstruction using Deconvolution \& Debluring}

Reconstruction of microscopic images has been crucial to improving the quality of the images as it is an essential task to reverse the effects that occur while capturing the images. One major contributor in optical microscopes is light diffraction, which needs to be restored, along with some other artifacts such as blur and low contrast.

Deconvolution is a computational process that mathematically tries to reassign the image that occurred due to light diffraction to its original supposed image.
\begin{equation}
    Y=K* \bigotimes X+N\label{eq}
\end{equation}

Deconvolution is all about estimating the $X$ from its degraded acquired $Y$ obtained by ``\eqref{eq}''
where $\bigotimes$ is the convolution operation, $K*$ denotes a blur kernel, and $N$ stands for noise. The blur kernel is often calculated using the point spread function (PSF), which is the mathematical calculation of the light diffraction of the imaging capture device. Deconvolution works on estimating the PSF and trying to reverse the diffraction to obtain the original image.


Traditionally, deconvolution has been achieved through methods like RL, Wiener filters, and blind deconvolution techniques like MLE. However, recent advancements have brought forward the use of DL to perform deconvolution beyond the mathematical limit. However, with the recent advancements in DL, new studies have been trying to incorporate it into conventional methods. For example, Chobola et al.\cite{chobola2023} developed a method to restore sharp images from the original blurred ones. Their novel method combines a classic iterative RL deconvolution algorithm and a pre-trained network to form \emph{CiDeR}. After the feature selection process, their use of the RL algorithm helped skip the large neural net, reducing the computational required without losing results. Thus proving it to be a valuable tool in image reconstruction. Although it may seem like DL require too much data to be trained on for better results, Saha showed through their paper\cite{saha2020} that even correctly generated synthetic data could be applied to real images acquired on different microscope modalities successfully. Their architecture uses a CNN model called \emph{ PHASENET}, which consists of five stack blocks of 3x3x3 conv layers and one max-pool layer followed by two dense layers. This model is then trained on synthetic data and tested on real data to produce better results than its predecessors with better efficiency. Vinogradova\cite{vinogradova2022} further developed this method to include even aberrations in the synthetic data to make the model capable of restoring precisely from distorted images compared to just point-wise images from before. Light-sheet Fluorescence Microscopy (LSFM)\cite{stelzer2021} has recently become more popular due to its long-term observation of live cells, thus garnering more use in various fields. Bai et al.\cite{bai2019} has developed a novel architecture for LSFM, that can reconstruct high-quality images using just a single scan of BB light-sheet microscopy. This model is not only cost-effective but also about 100 times faster. Recently, GANs have been gaining popularity in image reconstruction as GANs are designed to generate new images through adversarial training, which means they can create realistic-looking images based on distorted or incomplete images\cite{bhadra2020medical}. Lee et al.\cite{lee2019} developed a 3D blind image deconvolution network for FLM using a spatially constrained cycleGAN\cite{fu2018three}. They inference and trained their network (SpCycleGAN) in 3 axis - $xy, yz\ \&\ zy$ sections to incorporate all the 3D data, thus obtaining excellent results as compared to their predecessor algorithms, successfully being able to restore blurry and noisy volume to good quality volume so that deeper data can be used for the biological research. Meanwhile, Lim et al.\cite{lim2020} modified the cycleGAN\cite{fu2018three} with a traditional blur kernel in wide-field FLM, which can be used in blind and non-blind deconvolutions. They proposed a new penalized least squares (PLS) transportation cost function, that employs several benefits, particularly data consistency, to regularize the generator. Thus, it helps them to use a UNet block for the high-resolution image generator and a low-resolution image generator based on a 3D conv layer model with a 3D blur kernel. It also uses the original cycle GAN's multi-PatchGANs\cite{isola2017image}. Their model significantly increased over the traditional cycleGAN, specifically in non-blind cases.


Due to deconvolution often requiring high computations and the difficulties in obtaining accurate PSFs, alternate deblurring methods are also being explored extensively. Zhao et al.\cite{zhao2020} introduced a novel DL-based deblurring method that can be applied irrespective of the modality. They used a Residual dense network within the CNN, which consists of five blocks: shallow feature extraction, residual dense blocks (RDBs), global feature fusion, global residual learning, and up-scaling in a feed-forward fashion. The residual connections introduced in the RDB further improve the information propagation, thus having more parameters to learn. This model outperforms all three of their comparison criteria (MSE, PSNR, SSIM) and shows that their method improves the system's quality by 2.14x times. Motion blur also significantly affects microscopy images through causes like sample movement. Although deconvolution could fix the defocus blur, motion blur must be tackled differently. Although general regression neural network (GRNN)\cite{yan2016blind} and residual dense network (RDN)\cite{zhao2020new} are used to solve motion blur, Jiang et al.\cite{jiang2020} created a method that can automatically deblur a microscopic image regardless of whether it is defocus or motion blur type. Their model, called deep blind microscopic image deblurring (DBMID), employs a two-phase deblurring workflow starting with a VGG-16-based\cite{simonyan2015} classification network to categorize the image into one of the four pre-determined blur types - in focus, defocus, motion, and mixed blur. Then, depending on the type, it forwards it into either a defocus or motion blur DL-based network, which then outputs a final image. Their classification accuracy on testing data came to about 99.77\%. They also tested their model on various test data, validating its generalizability. However, DBMID has a significant limitation in that it takes considerable time for the deblurring process, negating it as an ideal solution for real-time imaging applications.

In blind deconvolution, Shajkofci et al.\cite{Shajkofci2020} improves the resolution of light microscopy images of non-flat objects by locally estimating image distortion while jointly estimating object distance to the focal plane. They estimated the parameters of a spatially-variant PSF model using a CNN, which does not require instrumentor object-specific calibration. Deconvolution methods used to deblur microscopy images have the major disadvantage of slow processing time and result in a sub-optimal performance if the operator chosen to represent PSF is not optimal. Employing a deep learning approach shows significant improvement over these disadvantages, and it certainly has the potential to work in a wide range of optical imaging systems\cite{zhao2020}.

\subsection{Denoising}

Noise has always been a part of imaging system, and solving this is one of the core image processing techniques. Compared to traditional Gaussian noise in photography, microscopy images are filled with Poisson noise due to the optical signal needing quantization\cite{zhang2019a}. Denoising can be reduced by having longer exposure times, which may photo-damage the sample, resulting in noisy images.

Using DL in denoising has brought many advancements, especially regarding optical microscopy. Zuluaga et al.\cite{gilzuluaga2021} aimed to introduce a lightweight CNN for medical image denoising. The architecture of the proposed model consisted of an inception reduction block with an encoder/decoder network, which was able to remove various levels of noise from the medical images, which have been shown to overcome state-of-the-art models in the same domain. However, the model is yet to be trained on different medical image typologies and higher degrees of noise map spatial distributions. To eliminate the need for clean target images and pairs of noisy images, NOISE2VOID (N2V) a novel image denoising training scheme that will require only a single noisy image to train denoising CNNs was introduced by Krull et al.\cite{krull2019}. N2V's performance when compared to traditional denoising methods, particularly in biomedical image data where obtaining such training targets is challenging, the average PSNR soared high. Performance evaluation on the BSD68 dataset and simulated microscopy data reveal N2V's competitive edge against traditionally trained networks, N2N-trained networks, and training-free denoising methods like BM3D\cite{yahya2020} and non-local means. N2V employs a blind spot network within the CSBDeep framework\cite{sharon} based on UNet architecture because of which N2V experiences a moderate performance drop compared to methods with more information during training, struggles with predicting pixels in images with high irregularities, and is vulnerable to structured noise violating pixel-wise independence assumptions.
To address the limitations of image denoising in existing methods like DCNNs and GANs. My Vo et al.\cite{vo2021} introduced BoostNet. This innovative approach integrates a deep convolutional neural network and a robust generative adversarial network into an ensemble, significantly enhancing denoising performance. Experiments on challenging datasets show BoostNet outperforms other denoisers in quantitative metrics and visual quality. Its potential applications span medical imaging, object recognition, and mobile robot operations, owing to its adeptness in managing noise while preserving essential image details. However, BoostNet's computational runtime could be faster due to its large number of trainable parameters. Tureyen et al.\cite{demircan-tureyen2022} proposed a novel approach to address the challenge of denoising fluorescence microscopy images CNN on a tailored dataset. They used a generalized Anscombe transform (GAT) to enhance the model capability. Their study demonstrated that their model could approximate fluorescence images effectively, even with a limited amount of actual fluorescence data. It showcases the potential of transfer learning from tailored datasets in microscopy image restoration. Since they focused on specific types of noise, i.e., Gaussian and MPG, the generalizability of the approach is limited. Also, using a limited amount of data for fine-tuning may pose challenges in fully capturing the variability when presenting real microscopy images.

\section{Conclusion}
This survey paper has explored the recent advancements in microscopy image enhancement using DL. The paper, through an in-depth analysis, highlighted significant contributions made through imaging techniques like SR, reconstruction using deconvolution and deblurring, and denoising using DL, which enabled higher image quality, enhanced analysis, which further enables higher research in the field of microscopy. The present study has studied some evolution of microscopic image enhancement over time, research gaps, challenges towards building advanced models, possible recommendations to overcome those challenges, and determining the best performance model.

The survey unveils the effectiveness of DL-based SR methods, which enhance spatial resolution and thus enable a detailed view of biological structures. Reconstructing HR images from a blurred, LR input, thus mitigating noise and blur, has always been a longstanding challenge in microscopy imaging. Such applications in reconstructing have demonstrated tremendous latent capabilities of dl. Moreover, DL-based denoising methods have also been effective, especially in enhancing SNR, which produces much better images and leads to precise interpretation of images. DL-based denoising can preserve crucial structural information, opening new avenues for advanced analysis and research in several scientific disciplines. The focus is on an exhaustive and comprehensive survey of deep learning-based research papers on microscopic image enhancement. The present work will benefit researchers by providing a clear direction for future scope in microscopic image enhancement and analysis.

\subsection{Future Direction}

After the introduction of the attention mechanism and ViT\cite{dosovitskiy2021}, a rapid advancement in the field of Image Processing followed. Getting on this trend, various work using transformers on image enhancement have been done, and promising results can be shown while bringing such methods onto microscopy image enhancement. Zhang et al.\cite{zhang2022d} proposed an attention retractable transformer for image restoration, showing remarkable results in image super-resolution, denoising, and artifact reduction tasks. Moreover, exploring graph neural networks for encapsulating the complex spatial relationships among microscopy images could lead to better interpretability and higher accuracy results\cite{kebaili2023,ma2020a}.

\IEEEtriggercmd{\enlargethispage{-5em}}

\bibliography{ref}

\end{document}